\documentclass[10pt,leqno]{article}

\usepackage{graphicx}
\usepackage{indentfirst,csquotes}

\usepackage{amssymb,amsthm,amsmath}
\usepackage{xcolor,paralist,hyperref,titlesec,fancyhdr,etoolbox}
\newtheorem{theorem}{Theorem}[]

\usepackage{lipsum}

\usepackage{multirow}
\usepackage{booktabs}
\usepackage{longtable}
\usepackage{pdflscape}
\usepackage{listings}
\usepackage{amssymb}
\usepackage{graphicx,subcaption}
\usepackage{caption}
\usepackage{amsmath}
\usepackage[linesnumbered,ruled,vlined]{algorithm2e}
\usepackage{footnote}
\makesavenoteenv{tabular}
\makesavenoteenv{table}
\usepackage{xcolor}

\def \Z {\mathbb{Z}}
\def \R {\mathbb{R}}

\def \Q {\mathbb{Q}}

\def \x {\mathbf{x}}
\def \y {\mathbf{y}}

\def \v {\mathbf{v}}
\def \n {\mathbf{n}}

\begin{document}
\title{A semi-algebraic model for automatic loop parallelization\thanks{This work is supported by the Chongqing Talents Plan Youth Top-Notch Project (2021000263). }}
%
%
\author{Changbo Chen\\
Chongqing Institute of Green and Intelligent Technology, \\
Chinese Academy of Sciences, Chongqing 400714, China 
}
%
%
%
\date{}
\maketitle              

\begin{abstract}
In this work, we introduce a semi-algebraic model 
for automatic parallelization of perfectly nested polynomial loops, 
which generalizes the classical polyhedral model. 
This model supports the basic tasks for automatic loop parallelization, such as the representation of the nested loop, the dependence analysis, the computation of valid schedules
as well as the transformation of the loop program with a valid schedule. 
\end{abstract}

\section{Introduction}

The polyhedral model~\cite{Feautrier2011} has become a standard model for automatically parallelizing affine nested loops. Pluto~\cite{bondhugula2008practical} and PPCG~\cite{2013Polyhedral} 
are representative compilers for automatic parallelization developed based on the polyhedral model targeting respectively on the CPU and GPU architecture.
Today, the polyhedral model has found important applications in both the field of scientific computing~\cite{Benabderrahmane2010} and deep learning~\cite{TensorComprehension2018}.
Although many loops in practice can be modeled in this framework, it remains theoretically interesting  to handle nonlinear loops not supported by such a model. 

There are already some attempts towards generalizing the polyhedral model for nonlinear loops.
One of the pioneering work is done by Armin Gr\"{o}\ss{}linger in his dissertation~\cite{Armin2010}
which discusses the handling of nonlinear expression in all phases of the parallelization process, including the dependence analysis, schedule computation and  code generation, with different techniques.
In particular, it presents in detail how to apply quantifier elimination for the dependence analysis and the computation of schedules and how to use cylindrical algebraic decomposition~\cite{col75}  for the code generation. 
A further development of this work is reported in~\cite{GROLINGER2006} focusing 
on the use of quantifier elimination.

Although we also use cylindrical algebraic decomposition as the main tool, our formulation is different from~\cite{Armin2010}. 
The main differences lie on the iteration variables and the representation 
of the access functions in the new loop. 
The iteration variables for the new loop may include the old iteration variables
and the access function may remain unchanged in~\cite{Armin2010}
while in four framework only new iteration variables are present
and the access functions are modified to be functions of them.

Another important direction on automatic parallelization of polynomial loops is proposed by Paul Feautrier~\cite{Feautrier2015},
which discusses the possibility of leveraging Positivstellensatz to handle polynomial constraints appearing in  dependence analysis and schedule computation. Such technique can be seen as a generalization 
of Farkas lemma for the linear case.

\section{Background on semi-algebraic sets}
In this section, we recall some basic notions and results on semi-algebraic sets and semi-algebraic functions. 
A comprehensive introduction of them can be found in the classical textbook~\cite{bpr06}.
Throughout this paper, let $\Z$ be the ring of integers,  $\Q$ be the field of rational numbers, and $\R$
be the field of real numbers. 

Let $F$ and $G$ be finite in $\Q[\x]$, where $\x=x_1,\ldots,x_s$.
We call the set $S:=\{\x\in \R^s \mid F(\x)=0, G(\x)>0\}$ a {\em basic semi-algebraic set}. 
The system $sas:=\{F(\x)=0, G(\x)>0\}$ is called a {\em basic semi-algebraic system}
and a {\em defining system} of $S$, denoted by $E(S)$. 
Its zero set, denoted by $Z(sas)$,  is $S$.
A set $S\subseteq \R^s$ is called a {\em semi-algebraic set} if it can be written as a finite union of basic semi-algebraic sets. Accordingly, there is a semi-algebraic system defining $S$, 
which is a sequence of basic semi-algebraic systems $sas_1,\ldots,sas_e$ 
such that $S=\cup_{i=1}^e Z(sas_i)$.
In practice, sometimes one could merge the basic semi-algebraic systems in a sequence 
into more compact forms without changing the zero set. 
For instance, two basic semi-algebraic system $\{f=0, g=0\}$ and $\{f=0, g>0\}$
could be merged as one system $\{f=0, p\geq 0\}$.

It is not hard to show that when applying the basic set-theoretical operations, namely the complement, the union and the intersection, on semi-algebraic sets finitely many times, the resulting sets are still semi-algebraic. 
Let $S\subseteq \R^s$ be a semi-algebraic set.
Let $\pi_k$, or $\pi_{\x}$, be the projection operation defined as $\pi_k(S)=\{\x\in\R^k\mid \exists \y\in \R^{s-k}, (\x,\y)\in S\}$.
It is well-known that $\pi_k(S)$ is also a semi-algebraic set. 

Let $S\subseteq \R^s$  and $T\subseteq \R^t$ be two semi-algebraic sets. A function $f:S\rightarrow T$ is called a {\em semi-algebraic function} if the graph of $f$, namely $S\times f(S)$ is a semi-algebraic subset of $\R^{s+t}$.
It is not hard to prove the following property: 
the image of a semi-algebraic function over a semi-algebraic set is semi-algebraic; the preimage of a semi-algebraic function over a semi-algebraic set is semi-algebraic; the composition of two semi-algebraic functions is semi-algebraic. 

Let $S\subseteq \R^s$. 
If $s=1$, a {\em cylindrical algebraic decomposition} (CAD) of $S$ is a family of finitely many semi-algebraic sets ${\cal C}=\{C_1,\ldots,C_e\}$ such that $S$ is a disjoint union of elements in ${\cal C}$. Moreover, each $C_i$ is either of the form $\gamma_i\in \R$ or of the form $(\alpha_i,\beta_i)$, where $\alpha_i<\beta_i$ and each $\alpha_i,\beta_i,\gamma_i$ is $-\infty$, $+\infty$, or a real algebraic number. 
For $s>1$, a CAD of $S$ is a family of finitely many non-empty semi-algebraic sets ${\cal C}=\{C_1,\ldots,C_e\}$ such that $S$ is a disjoint union of them. Moreover, the following properties are required:
\begin{itemize}
     \item For any $i\neq j$, we have $\pi_{s-1}(C_i)=\pi_{s-1}(C_j)$ or $\pi_{s-1}(C_i)\cap \pi_{s-1}(C_j)=\emptyset$.
    \item  The set ${\cal C}'=\{\pi_{s-1}(C_1), \ldots, \pi_{s-1}(C_e)\}$, where identical elements have been removed, is a CAD of $\pi_{s-1}(S)$. 
    \item On each cell $C'_i$ of ${\cal C}'$, there are finitely many continuous semi-algebraic functions $\xi_{i,1}<\cdots<\xi_{i,e_i}$ defined on $C'_i$ such that any cell $C_{i}$ satisfying $\pi_{s-1}(C_{i})=C'_i$ is in one of the following forms: $\{(\x,y)\in \R^s\mid \x\in C'_i, y=\xi_{i,k}(\x)\}$, $\{(\x,y)\in \R^s\mid \x\in C'_i, \xi_{i,k}(\x)<y<\xi_{i,k+1}(\x)\}$, $\{(\x,y)\in \R^s\mid \x\in C'_i,-\infty<y<\xi_{i,1}(\x)\}$, or $\{(\x,y)\in \R^s\mid \x\in C'_i,\xi_{i,e_i}(\x)<y<+\infty\}$.
        
\end{itemize}
It is easy to show that the set $\pi_i({\cal C})=\{\pi_{i}(C_1), \ldots, \pi_{i}(C_e)\}$, $1\leq i\leq s$, where identical elements have been removed, is a CAD of $\pi_{i}(S)$, called the {\em induced CAD} of $S$ in $\R^i$. 
In the above definition of CAD, the order of cells in the CAD is not specified. 
But one can assign an order to the cells in a natural way.
For any two cells $C_{i_1}$ and $C_{i_2}$ in the CAD ${\cal C}$, we say $C_{i_1}\prec C_{i_2}$ if 
there exists a $k$, $1\leq k\leq s$, such that $\pi_{k-1}(C_{i_1})=\pi_{k-1}(C_{i_2})$ and two points $p=(x_1,\ldots,y_s)\in C_{i_1}$ and $q=(x_1,\ldots,y_s)\in C_{i_2}$ satisfying that $x_k<y_k$ and $x_i=y_i$, for all $1\leq i<k$.
In practice, it is convenient  to organize the cells in a CAD ${\cal C}$ of $\R^s$ in a tree $T$, called {\em CAD tree}, which can be constructed recursively as follows:
\begin{itemize}
    \item There is a root node $r$. Its children are exactly the elements of $\pi_1({\cal C})$. The children are sorted from left to right by the order $\prec$.
    \item Let $T'$ be the CAD tree corresponding to $\pi_{s-1}({\cal C})$.
    \item For each leaf $C'$ of $T'$, its children are exactly the elements of ${\cal C}$
    whose projections on $\R^{s-1}$ are the same as $C'$. The children are sorted from left to right by the order  $\prec$.
\end{itemize}
The zero set of this tree is defined as the union of all CAD cells in it, denoted dy $Z(T)$.




\section{An abstract model for loop programs}
In this section, we define the type of loops that we target on in an abstract way. 
Briefly, we require that the loops can be organized in a sequence of trees.
Depending on the functions involved in the loop bounds, access functions, 
and schedules being affine or nonlinear, the loop can be studied 
in a polyhedral model or a semi-algebraic model. 

A perfectly nested program is a sequence $P, L_1,\ldots, L_s, stmt$.
Here $P\subseteq\R^p$ and $P_z := P\cap \Z^p$ defines the ranges of the program parameters ${\bf n}=n_1,\ldots,n_p$. 
Each $L_i=(x_i, I_i)$ represents a loop, where $x_i$ denotes the iteration variable, $I_i$
is an interval representing the loop bounds, and the variable $x_i$ takes values in $I_i\cap\Z$.
Let $\ell_i$ and $u_i$ be the left and right endpoints of $I_i$. 
They are continuous functions of ${\bf n}, x_1,\ldots,x_{i-1}$
if $({\bf n}, x_1,\ldots,x_{i-1})$ takes values in $P\times I_1\times\cdots\times I_{i-1}\subseteq\R^{p+i-1}$. 
A statement $stmt$ is a quadruple $(\x, A_0,{\bf A}, {\bf R})$, where 
$\x=x_1,\ldots,x_s$ are the iteration variables, $A_0$ is a multi-dimensional array (or tensor), ${\bf A}=f(A_1,\ldots,A_r)$ is a function of tensors $A_1,\ldots,A_r$, and ${\bf R}$ is a reduction/aggregation operation, 
including the assignment operation being a special case.
The indices  of each $A_j$, $j=0,\ldots,r$,  are  functions of  ${\x}$ and are denoted by $g_0({\x}), \ldots, g_r({\x})$. They are called {\em access functions} of $stmt$. 
The statement $stmt$ gathers all the values of ${\bf A}$ corresponding to the same indices $g_0({\x})$
of $A_0$, applies the reduction operation ${\bf R}$, and writes to $A_0[g_0({\x})]$.
For any fixed ${\bf n}\in P$,
let $D:=I_1\times\cdots\times I_s$
and $D_z := D\cap \Z^s$. 
The set $D_z$ gives the iteration domain of the statement. 
Any point of $D_z$ is called an iteration vector.

Let $P$ be defined as before.
A nested loop program is a tree rooted at $P$ such that each path is a perfectly nested program. 
A loop program is a sequence of nested loop programs. 
Thus a loop program is a forest, where each tree denotes a nested loop program. 

Let $L$ be a loop program. Let $stmt'(\x')$ and $stmt''(\x'')$ be two different statement instances. If  two instances access the same memory address, and one of them is write, we say that there is a {\em data dependence} in the loop program $L$.  
Let us illustrate this notion for the case that $stmt'$ and $stmt''$ are the same statement $stmt$ belonging to the same nested loop $P, L_1,\ldots,L_s, stmt$. 
For any given ${\bf n}\in P_z$,
this is equivalent to saying that there exist a $i=0,\ldots, r$ and two iteration vectors ${\bf x}'\neq {\bf x}''$
such that $A_0[g_0({\x}')]=A_i[g_i({\x}'')]$.
That is $A_i=A_0$ and $g_0({\x}')=g_i({\x}'')$ must simultaneously hold.
In general, one could build a {\em dependence graph}, where each node denotes 
a statement. One adds a directed edge $(stmt', stmt'')$ to the graph 
for any  data dependence between an instance of $stmt'$ and an instance of $stmt''$, 
where the instance of $stmt''$ is executed after that of $stmt'$.
Note that there could be multiple edges between two nodes.
For each edge, there is an associated {\em dependence set}, denoted by $DS$ to describe the dependence.

To illustrate the concept of $DS$, let us consider a perfectly nested loop $P, L_1,\ldots,L_s, stmt$. 
Here $stmt=(\x, A_0, f(A_1,\ldots,A_r), {\bf R})$. 
 Then there are three possible dependences (RAW, WAW, WAR) between $A_0[g_0]$ and itself. 
If there is an $i=1,\ldots, r$, such that $A_0=A_i$. 
 Then there are also three possible dependences (RAW, WAW, WAR) between $A_0[g_0]$ and $A_i[g_i]$. 
For a fixed $i$.
Let $DS := \{({\bf n}, {\bf x}', {\bf x}'')\in\R^{p+2s}\mid {\x}'\prec {\x}'', {\bf n}\in P, {\x}'\in D, {\x}''\in D, g_0({\x}')=g_i({\x}'')\}$.
All three dependences correspond exactly to the same dependence set $DS_z:=DS\cap\Z^{p+2s}$.
Namely, $DS_z = \{({\bf n}, {\bf x}', {\bf x}'')\in\Z^{p+2s}\mid {\x}'\prec {\x}'', {\bf n}\in P_z, {\x}'\in D_z, {\x}''\in D_z, g_0({\x}')=g_i({\x}'')\}$.

Given a statement $stmt$, let ${\bf x}=(x_1,\ldots,x_s)$ be its iteration variables.
A schedule $sch$ for the statement is a multidimensional function 
${\bf y}=M({\bf x})$, whose domain is the iteration domain of $stmt$. 
A schedule is valid if it maintains in the new iteration space ${\bf y}$ 
 the execution order of the instances of the statement whose corresponding iteration vectors satisfy the dependences .  
That is, we need to solve the following problem:
\begin{equation}
\label{eqs:sch1}
    \forall DS_z, \forall ({\bf n}, {\x}', {\x}'')\in DS_z, M({\x}')\prec M({\x}'').
\end{equation}

\section{The semi-algebraic model for a loop program}
In this section, we introduce the semi-algebraic model for the type of loop programs
presented in last section.

\subsection{The loop program associated with a CAD}

Let $\x\in\R^s$ and ${\bf n}\in\R^p$ be two disjoint set of variables.
Fix the variable ordering ${\bf n}\prec\x$.
We write also $\y=({\bf n}, \x)=(y_1,\ldots,y_p,y_{p+1},\ldots,y_{p+s})$.
Let $S\subseteq\R^{p+s}$ be a semi-algebraic set. 
Let ${\cal C}$ be a CAD of $S$. 
Let $C\in{\cal C}$ be a cell of ${\cal C}$. 
We can always write $C$ as a Cartesian product  $C=C_1\times\cdots \times C_p\times C_{p+1}\times \cdots \times C_{p+s}$, 
where each $C_i$, $i=1,\ldots, p+s$ is of the form $y_i=\xi_i(y_1,\ldots,y_{i-1})$ or $\ell_i(y_1,\ldots,y_{i-1})<y_i<u_i(y_1,\ldots,y_{i-1})$.
Let $P := \left(C_1\times\cdots\times C_p\right)$.
For $p+1\leq i\leq p+s$, if $C_i$ is of the form $y_i=\xi_i(y_1,\ldots,y_{i-1})$, let $L_{i-p}=(x_{i-p}, [\xi_i, \xi_i])$. If $C_i$ is of the form $\ell_i(y_1,\ldots,y_{i-1})<y_i<u_i(y_1,\ldots,y_{i-1})$, let $L_{i-p}=(x_{i-p}, ( \ell_i,  u_i) )$. 
Let $stmt=(\x, A_0[g_0({\bf n}, \x)], f(A_1[g_1({\bf n}, \x)],\ldots,A_r[g_r({\bf n}, \x)]), {\bf R})$ be a statement,
where the access functions $g_i$, $i=0,\ldots,r$,
are continuous semi-algebraic functions.
We call $P, L_1,\ldots,L_s, stmt$ a  perfectly nested loop associated with the CAD cell $C$.

Let ${\cal C}$ be a CAD tree of $S$.
Let ${\cal P}=\pi_{p}({\cal C})$ be the induced CAD tree of ${\cal C}$.
For any leaf $P$ of ${\cal P}$, all its descendants form a subtree of ${\cal C}$ rooted at $P$. 
For any CAD cell $C$ of ${\cal C}$ such that $\pi_p(C)=P$ and a statement with the iteration variables ${\bf x}$, 
we can associate with $C$ a perfectly nested loop as before. 
As a result, we obtain a nested loop program rooted at $P$, 
which is called an {\em associated  nested loop program} with the tree rooted at $P$.
The loop program formed by all these nested loop programs is called 
an {\em associated loop program} with ${\cal C}$. 

\subsection{A perfectly nested polynomial loop}
In this section, we introduce a class of perfectly nested polynomial loops, 
for which we demonstrate how to conduct dependence analysis, compute a valid schedule, 
and transform the initial loop with a schedule.

Let ${\cal L}:=P, L_1,\ldots,L_s, stmt$ be a perfectly nested loop.
Throughout this section, we make the following assumptions. 
The set $P$ is a semi-algebraic set of $\R^p$.
It is defined by a semi-algebraic system $E(P)$.
For each $L_i=(x_i, I_i)$, the endpoints of $I_i$ are polynomial functions.
The interval $I_i$ naturally defines a semi-algebraic system $E_i:=\{x_i~\sigma_{\ell}~\ell_i, x_i~\sigma_{u}~u_i\}$, 
where $\sigma_{\ell}$ is $>$ (resp. $\geq$) if $I_i$ is open (closed) on the left, 
and $\sigma_{\ell}$ is $<$ (resp. $\leq$) if $I_i$ is open (closed) on the right.
Let $stmt=(\x, A_0[g_0({\bf n}, \x)], f(A_1[g_1({\bf n}, \x)],\ldots,A_r[g_r({\bf n}, \x)]), {\bf R})$. The access functions $g_i$, $i=0,\ldots,r$ are polynomial functions.
Under these assumptions, the loop ${\cal L}$ is called a {\em perfectly nested polynomial loop}.
Recall that, for any fixed ${\bf n}\in P$, we have $D:=I_1\times\cdots\times I_s$
Let $E_D:=\cup_{i=1}^s E_i$. 
Assume that $E(P)$ is given by a sequence of basic semi-algebraic systems $sas_1,\ldots,sas_e$. 
The notation $E(P) \cup E_D$ means the sequence $sas_1\cup E_D,\ldots, sas_e\cup E_D$.
Let $E_{\cal L}:=E(P) \cup E_D$. 
It  is a semi-algebraic system and we have $Z(E_{\cal L})=P\times D$ and $Z(E_{\cal L})\cap\Z^{p+s}=P_z\times D_z$.

\subsubsection{Dependence analysis}
For a dependence between $A_0[g_0]$ and $A_i[g_i]$, where $A_0=A_i$, 
let $DS := \{({\bf n}, {\bf x}', {\bf x}'')\in\R^{p+2s}\mid {\x}'\prec {\x}'', {\bf n}\in P, {\x}'\in D, {\x}''\in D, g_0({\x}')=g_i({\x}'')\}$ and $DS_z:=DS\cap\Z^{p+2s}$.
Since $DS$ is a semi-algebraic set, its emptiness is decidable. 
If $DS$ is empty, there is no true dependence between  $A_0[g_0]$ and $A_i[g_i]$. 
If $DS$ is not empty, it is still possible that $DS_z$ is empty. 
Note that for a fixed value of $n$, the set $DS_z$ is finite, thus the emptiness of $DS_z$
is still decidable. 

\subsubsection{Computation of the schedule}
We consider a multi-dimensional schedule $M=(M_1,\ldots,M_s)$, where each $M_i$ is a function of ${\bf n}$ and $\x$. 
The schedule is valid if and only if Eq.~(\ref{eqs:sch1}) holds.
For a fixed value $n$, as $DS_z$ is finite, this is decidable. 
Now suppose that the schedule $M$ is parametric, we could like to find the conditions on the parameters to make $M$ valid. 
Again if the values of the parameters are only allowed to be taken from a finite set, this problem 
is still decidable. 
Without such restriction, it may not be decidable. 
One could then try to solve the following real quantifier elimination problem and obtain the conditions on the parameters to make $M$ valid. 
\begin{equation}
\label{eqs:sch2}
    \forall DS, \forall ({\bf n}, {\x}', {\x}'')\in DS, M({\x}')\prec M({\x}'').
\end{equation}
Obviously, if Eq.~(\ref{eqs:sch2}) holds, then Eq.~(\ref{eqs:sch1}) also holds.
In general, one may would like to maximize the parallelism while making sure that $ M({\x}')\prec M({\x}'')$.
To achieve this, one could first try to search schedules such that $M_i(\x')=M_i(\x'')$
for all $1\leq i\leq s-1$ and $M_s(\x')<M_s(\x'')$. 
If this fails, we could  try to search schedules such that $M_i(\x')=M_i(\x'')$
for all $1\leq i\leq s-2$ and $M_{s-1}(\x')<M_{s-1}(\x'')$, and so on and so forth. 
\subsection{Transformation of the loop program}
Solving Eq.~(\ref{eqs:sch2}) brings us a semi-algebraic system $E_V$  
on the parameters $\v$ of the schedule $M$  of  a perfectly nested loop 
${\cal L}:=P, L_1,\ldots,L_s, stmt$. 
Suppose that the zero set of $E_V$ is not empty, 
then any point of it gives a valid schedule that maximizes the parallelism.
We could apply $M$ to the iteration space $\x$ of ${\cal L}$ to obtain the new iteration space $\y$, 
for which the parallelism is explicit. 
To do so, we need to solve the semi-algebraic system $E_T$
below. 
Let $E_M:=\{y_1-M_1(\x), \ldots, y_s-M_s(\x)\}$.
Let $E_T := E_{\cal L}\cup E_M\cup E_V$.
Now the variables appearing in $E_T$ are ${\bf v}, {\bf n}$, $\x$ and $\y$. 
We consider a new variable ordering $ord:={\bf v}\prec {\bf n}\prec\y\prec\x$. 
We compute a CAD ${\cal C}$ of the zero set of $E_T$ under the variable ordering $ord$.
For each cell $C(\v)$ of $\pi_{\v}({\cal C})$ and  the tree rooted at $C(\v)$, 
if for each path $\Gamma$ of the tree, each node is of the form $x_i=\xi_i(\v, \n, \y)$, 
we keep the cells of ${\cal C}$ with $C(\v)$ as ancestors and name the resulting CAD tree as $T_{\cal L}$.

Now we are ready to make the transformation. 
For a given cell $C(\v)$, all its descendants form a subtree  of $T_{vnyx}$ rooted at  $C(\v)$.
For any given value of  $C(\v)$, this subtree defines a CAD tree $T_{nyx}$ in the space of $\n$, $\y$ and $\x$.
This tree has a special shape, its projection onto the space of $\n$ and $\y$  is still a CAD tree, 
named as $T_{ny}$. Moreover, for each leaf of it, the descendants of the leaf form a single path in   $T_{nyx}$ 
of the shape $x_1=\xi_1(\n, \y), \ldots, x_s=\xi_s(\n, \y)$, written as $\x=\xi(\n,\y)$ for short. 
Recall that the input perfectly nested loop is $P,L_1,\ldots,L_s,stmt$, 
where $stmt=(\x, A_0[g_0({\bf n}, \x)], f(A_1[g_1({\bf n}, \x)],\ldots,A_r[g_r({\bf n}, \x)]), {\bf R})$. 
Thus this loop will be transformed into a loop program associated with the CAD tree $T_{ny}$, 
where the statement attached to each leaf is 
$$
stmt'=(\y, A_0[g_0({\bf n}, \xi(\n,\y))], f(A_1[g_1({\bf n},\xi(\n,\y))],\ldots,A_r[g_r({\bf n}, \xi(\n,\y))]), {\bf R}).
$$

In practice, one often need to merge the CAD cells to 
reduce the number of loops in the resulting transformed loop program, for instance with the techniques in~\cite{DBLP:conf/casc/ChenM15}.

\begin{theorem}
\label{thm:real}
Let ${\cal L}:=P, L_1,\ldots,L_s, stmt$ be a perfectly nested loop.
Let $\y=M(\x, \v, \n)$ be a parametric schedule of ${\cal L}$, 
which is valid when the parameters take the value from $E_V$.
Let $T_{\cal L}$ be the CAD tree resulted from applying $M$ to ${\cal L}$
such that each $x_i$-node is of the form $x_i=\xi_i(\v, \n, \y)$.
 Suppose that the CAD tree $T_{\cal L}$ is not empty. 
For any cell $C(\v)$ of $\pi_{\v}(T_{\cal L})$ and any point of $C(\v)$, 
evaluating $T_{\cal L}$ at this point brings a CAD tree $T_{nyx}$, 
whose projection on the space of $\n$ and $\y$ is the CAD tree $T_{ny}$.
Then the following properties hold:
 \begin{itemize}
 \item[$(1)$]
 There is a one-to-one correspondence between the points of $Z(E_{\cal L})=P\times D$
 and the points of $Z(T_{ny})$.
 That is any point $(\n', \x')$ of $Z(E_{\cal L})$ is uniquely 
 associated with a point $(\n', \y')$ of $Z(T_{ny})$ and vice versa.
     \item[$(2)$] Let $(\n', \x', \x'')$ be a point in the dependence set $DS$
     and let $\y'=M(\n', \x')$ and  $\y''=M(\n', \x'')$. 
     Then $\y'\prec \y''$ holds. 
     Let $stmt_1$ and $stmt_2$ be respectively the statement corresponding to $\y'$ and $\y''$. 
     Then the statement instance $stmt_2(\y'')$ is executed after the statement instance $stmt_1(\y')$.
 \end{itemize}
\end{theorem}
\begin{proof}
It suffices to show that for any $\v'\in C(\v)$,  the map $M$ is a bijective from $E_{\cal L}$ to $Z(T_{ny})$. 
For any $\v'\in C(\v)$ and any point $(\n', \x')$ of $Z(E_{\cal L})$, let $\y'=M(\v', \n', \x')$.
As $E_T= E_{\cal L}\cup E_M\cup E_V$ and $C(\v)\subseteq Z(E_V)$, we have $(\v', \n', \y', \x')\in Z(E_T)$.
Since  $Z(E_T(\v'))=Z(T_{nyx})$, we have $(\n', \y', \x')\in Z(T_{nyx})$ and $(\n', \y')\in Z(T_{ny})$.  
Similarly, for any $\v'\in C(\v)$ and any $(\n', \y')\in Z(T_{ny})$, 
there exists $\x'$ such that $(\n', \y', \x')\in Z(T_{nyx})$. 
As $Z(E_T(\v'))=Z(T_{nyx})$, we have $\y'=M(\v', \n', \x')$ and $(\n', \x')\in Z(E_{\cal L})$.
Next we show that the map is injective. 
For any two points $(\n', \x')\neq (\n'',\x'')$ of $Z(E_{\cal L})$, 
let $\y'=M(\n', \x')$ and $\y''=M(\n'', \x'')$. 
We claim that $(\n', \y')\neq (\n'',\y'')$. 
Otherwise, as $x_i=\xi_i(\v, \n, \y)$, we must have $\n'=\n''$ and $\x'=\x''$, contradiction.

 

 If $(\n', \x', \x'')$ belongs to a dependence set $DS$, 
 we  have $\x'\prec\x''$.
 As the schedule $M$
 is valid for any $\v'\in C(\v)$, we must have $\y'\prec \y''$.
 Now consider the loop program associated with the CAD tree $T_{nyx}$. 
 As  $\y'\prec \y''$, let $k$ be smallest integer such that $y'_1=y''_1,\ldots,y'_{k-1}=y''_{k-1}$ and $y'_k<y''_k$. By definition, the first $k-1$ nested loop 
 of $stmt_1(\y')$ and $stmt_2(\y'')$ must be the same. 
 As $y'_k<y''_k$, either $stmt_1(\y')$ and $stmt_2(\y'')$ share the same $k$-th nested loop and $stmt_2(\y'')$ is executed after $stmt_1(\y')$, or the $k$-the nested loop of $stmt_2(\y'')$ is associated to a node right 
 to the one associated with $stmt_1(\y')$. 
 For both cases, $(3)$ holds.
\end{proof}

In Theorem~\ref{thm:real}, 
we demonstrate a valid transformation for the real case. 
Next theorem shows when a transformation is valid for the integer case. 

\begin{theorem}
    \label{thm:int}
Assume that $Z(T_{\cal L})\cap\Z^{|\v|+p+2s}\neq\emptyset$. 
Let $\v'\in\pi_{\v}(Z(T_{\cal L})\cap\Z^{|\v|+p+2s})$.
Assume that for any $(\n',\x')\in P_z\times D_z$,
$M(\v', \n', \x')\in \Z^{s}$.
We further assume that any function $\xi_i(\v, \n, \y)$ appearing in $T_{\cal}$ maps any integer point in the domain of the function to an integer.
Then the two properties of Theorem~\ref{thm:real} also hold
for the integer case.
\end{theorem}
\begin{proof}
 For any $\v'\in\pi_{\v}(Z(T_{\cal L})\cap\Z^{|\v|+p+2s})$
and any point $(\n',\x')$ of $P_z\times D_z$, 
as $\y'=M(\v', \n', \x')\in \Z^{s}$, 
by Theorem~\ref{thm:real}, 
we know that $(\v', \n', \y', \x')\in Z(T_{\cal L})\cap\Z^{|\v|+p+2s}$ and $(\n', \y')\in Z(T_{ny})\cap\Z^{p+s}$. 
Similarly, for any $(\n', \y')\in Z(T_{ny})\cap\Z^{p+s}$, 
the property of the function $x_i=\xi_i(\v, \n, \y)$ implies that $\x'=(\xi_1(\v', \n', \y'), \ldots, \xi_s(\v', \n', \y'))\in\Z^s$. 
Then the rest follows from the proof of Theorem~\ref{thm:real}.
\end{proof}




\section{Conclusion}
In this work, we have presented a theoretical 
model for automatic parallelization of polynomial loop programs. 
To make it practical, it is important to explore special structures 
appearing in various semi-algebraic systems derived from the model.
We have made an initial try in~\cite{DBLP:conf/cascon/ChenCKMX15}
to optimize our general algorithm for cylindrical algebraic decomposition and quantifier elimination~\cite{DBLP:journals/jsc/ChenM16},
but further optimizations are necessary.

%
%
%
\bibliographystyle{plain}

\end{document}